\documentstyle[aps,epsf,floats]{revtex}
\newcommand{\be}{\begin{equation}}
\newcommand{\ee}{\end{equation}}
\newcommand{\bea}{\begin{eqnarray}}
\newcommand{\eea}{\end{eqnarray}}

\begin{document}
\draft

\title{ Large Deviation Function   \\ 
		of the Partially Asymmetric Exclusion Process}

\author{ Deok-Sun Lee and Doochul Kim}

\address{
Department of Physics, Seoul National University  \\
Seoul 151-742, Korea}

\maketitle

\begin{abstract}
The large deviation function obtained recently by Derrida and Lebowitz for
the totally asymmetric exclusion process is generalized to the partially
asymmetric case in the scaling limit. 
The asymmetry parameter rescales the scaling variable in a simple way.
The finite-size corrections to the universal scaling function 
and the universal cumulant ratio are also obtained to the leading order.

\end{abstract}

\section{introduction}

The asymmetric simple exclusion process (ASEP) is the simplest
driven diffusive system where particles on a one-dimensional lattice
hop with asymmetric rates under excluded volume constraints.
Due to its simple but non-trivial out-of-equilibrium properties, 
it has attracted much attention recently. 
We refer to  \cite{derr} for a review of recent developments. 

For the prototype case of single-species, sequential updating dynamics, 
the time evolution operator of the probability distribution of particle
configurations turns out to be the asymmetric XXZ chain \cite{gwaspo,dkim}.
The latter  admits the Bethe ansatz solution for its eigenfunctions 
and eigenvalues when it is on a periodic ring. 
Due to its integrability,  one can obtain 
many exact results of physical interest.
In particular, the large deviation function (LDF) 
which describes the distribution of the total current 
has been obtained recently for a ring of $N$ sites with $P$ particles 
under a periodic boundary condition \cite{derrleb,derrapp}.
The LDF also describes the height distribution of the
Kardar-Parisi-Zhang (KPZ)-type growth models and is believed to be universal.
To confirm the universality of LDF,
Derrida and Appert \cite{derrapp} compared a cumulant ratio obtained from 
the analytic LDF  with numerical simulations of several stochastic models 
believed to belong to the KPZ universality class.

Since the LDF has been obtained in \cite{derrleb,derrapp} for 
the totally asymmetric exclusion process (TASEP) where particle hopping
occurs only to the right, 
it would be desirable to calculate it for 
the partially asymmetric exclusion process (PASEP) where
the particles can hop both to the right and to the left 
but with different rates.
In this paper, we report on this generalization using the crossover
scaling functions of the XXZ chain obtained previously in \cite{dkim}.
Our method assumes from the outset that $N$ is sufficiently large, 
but allows systematic evaluation of the finite-size corrections.
We reproduce the universal scaling function of \cite{derrleb,derrapp} 
for the PASEP and find that the asymmetry parameter rescales 
the scaling variable in a simple way. 
We also evaluate the leading order finite-size corrections to the universal 
scaling function and the cumulant ratio.

This paper is organized as follows. 
In Sec.~\ref{sec:large}, we introduce the model and notation.
In Sec.~\ref{sec:scaling}, we make the connection between the present problem
and the results of \cite{dkim} and derive the LDF for the PASEP. 
The finite-size corrections are evaluated in Sec.~\ref{sec:finite}. 
Sec.~\ref{sec:discussion} contains the summary and discussions, while 
Appendix shows the equivalence of two representations of the crossover 
scaling functions.

\section{Model and the Large Deviation Function }
\label{sec:large}

We consider the dynamics of the one-dimensional model in a periodic lattice
(ring) of $N$ sites with $P$ particles \cite{gwaspo}.
Each site $j$ ($1\leq j \leq N$) is either occupied by a particle
($\sigma_j=-1$) or vacant ($\sigma_j=+1$). 
The PASEP considered in this work
is defined by the following random sequential updating rule:
During each time interval $dt$, each particle can hop to its right or left
with probability $\frac{1}{2} (1 + \epsilon ) dt $ 
and $\frac{1}{2} (1 - \epsilon ) dt$, respectively, 
provided the target site is empty. 
$\epsilon$=1 corresponds to the TASEP considered in \cite{derrleb,derrapp} 
and we work in the region $0<\epsilon\leq1$.
Interpreting $\sigma_j$=$\pm1$ as the local slope of an interface in 
($1+1$) dimensions, 
one can map the model to the single step model \cite{derrapp,plischke}, 
an archetype of the KPZ-class models.
The quantity of main interest in this work is the total displacement $Y_t$ which
is the total number of hops of all particles 
to the right minus that to the left
between time 0 and time $t$.
In the single step model language, $Y_t$ is the total number of particles
deposited between time 0 and time $t$.

Let $ \sigma $ denote a system configuration $\{ \sigma_1, \ldots,\sigma_N \}$
and $P_{t} (\sigma) $ the probability of finding the system in
a configuration $\sigma$ at time $t$.
The master equation for the time evolution of $P_{t}(\sigma)$ can then
be written as
\be
\frac{dP_{t}(\sigma)}{dt} = -\sum_{\sigma'}
\langle \sigma | H | \sigma' \rangle P_{t}(\sigma'),
\label{masterP}
\ee
where $\langle \sigma | H | \sigma' \rangle$ is the representation, on the
basis where $\sigma_j^z$ are diagonal, of the time evolution operator $H$
given by
\be
H=- \sum_{j=1}^{N} \{(\frac{1+\epsilon}{2})\sigma_{j}^{+}\sigma_{j+1}^{-} + 
(\frac{1-\epsilon}{2})\sigma_{j}^{-}\sigma_{j+1}^{+} +
\frac{1}{4} (\sigma_{j}^{z}\sigma_{j+1}^{z} -1 )\}.
\ee
Here, $\sigma_j^{\pm}$ and $\sigma_j^z$ are the Pauli spin operators and
$\sigma_j$=$\pm 1$ are the eigenvalues of $\sigma_j^z$.

Next, following \cite{derrleb,derrapp}, we introduce $P_t(\sigma, Y)$, the
joint probability that the system is in a configuration $\sigma$ and $Y_t$,
the total displacement, takes the value $Y$ at time $t$, and let
\be
F_t(\sigma; \alpha) = \sum_{Y = -\infty}^{\infty} e^{\alpha Y} P_t(\sigma,Y).
\ee
Then, $F_t (\sigma; \alpha)$ evolves according to
\be
\frac{dF_t(\sigma; \alpha)}{dt} = \sum_{\sigma'} \langle \sigma | M | \sigma' 
\rangle F_t(\sigma'; \alpha),
\label{masterF}
\ee
where
\be
\label{m}
M = \sum_{j=1}^N \left\{ e^{\alpha} \left( \frac{1 + \epsilon}{2} \right)
\sigma_j^+ \sigma_{j+1}^- + e^{-\alpha} \left( \frac{1 - \epsilon}{2} 
\right) \sigma_j^- \sigma_{j+1}^+ + {1 \over 4 }  
(\sigma_j^z \sigma_{j+1}^z -1 ) \right\}.
\ee
The ``Hamiltonian" $-M$ is the asymmetric XXZ chain Hamiltonian studied, 
 e.g., in \cite{dkim}.
Let $\lambda (\alpha)$ denote the largest eigenvalue of $M$, regarded
as a function of $\alpha$. Then, one can show that
\be
\langle e^{\alpha Y_t} \rangle = \sum_{\sigma} F_t (\sigma; \alpha) 
\sim e^{ \lambda(\alpha) t} 
\label{Ylambda}
\ee 
as $t \rightarrow \infty$ and the long time behaviors of  
all cumulants of $Y_t$ are derived 
from $\lambda(\alpha)$.

The LDF $f$ describes the long time behavior of the distribution of $Y_t /t$
and is defined by
\be
f(y) = \lim_{t \rightarrow \infty} \frac{1}{t} \ln
\left\{ {\rm Prob} \left[ {Y_t \over t} = \bar{v}  + y  \right] \right\},
\label{eq:LDF}
\ee
where $\bar{v} = \lim_{ t \rightarrow \infty } \langle Y_t \rangle /t$ 
is the mean current for a ring of finite size $N$.
Note that 
$\bar{v} = \left. d \lambda(\alpha) / d \alpha \right|_{\alpha = 0}$. 
This can be easily obtained from a first order perturbation calculation as
\be
\bar{v}= \epsilon \rho (1- \rho) N { N \over N-1}.
\label{eq:exactv}
\ee
Our definition of $f(y)$ is slightly different from 
that of \cite{derrleb,derrapp} in that we use the exact value of $\bar{v}$, 
Eq.~(\ref{eq:exactv}), in Eq.~(\ref{eq:LDF}) 
while \cite{derrleb,derrapp} use its bulk value $\epsilon \rho (1 - \rho) N$.
Since 
$\langle e^{\alpha Y_t} \rangle \sim e^{ \lambda(\alpha) t}$  
on the one hand, and 
$\langle e^{\alpha Y_t} \rangle = \sum_{Y=-\infty}^{\infty} 
{\rm Prob} [ Y_t = Y ] e^{\alpha Y} \sim \max_y  e^{t(f(y)+\alpha \bar{v}
+ \alpha y)} $  on the other, 
the LDF is related to $\lambda(\alpha) - \alpha \bar{v} $ by the
Legendre transformation
\bea
f(y) &=& [ \lambda(\alpha)- \alpha \bar{v}] - \alpha y  ,  \label{leg_f}   \\
y &=& { d \over d \alpha } [ \lambda (\alpha) - \alpha \bar{v} ].  
\label{leg_y}
\eea
Therefore, the largest eigenvalue $\lambda(\alpha)$ of the asymmetric
XXZ chain $M$ determines the LDF. 

\section{$\lambda(\alpha)$ in the scaling limit}
\label{sec:scaling}

In \cite{derrleb,derrapp}, $\lambda(\alpha)$ for the case of $\epsilon$=1
is obtained for arbitrary $N$ and $P$. Then, one takes  the scaling limit,
$N\rightarrow \infty$, $\alpha \rightarrow 0$, with the scaling variable
$\alpha N^{3/2}$ and the density $\rho \equiv P/N$ fixed. In this scaling
limit,  $\lambda(\alpha)$ takes the parametric form
\be
\lambda(\alpha) = \alpha N \rho(1-\rho) +\sqrt{\frac{\rho(1-\rho)}{2 \pi
 N^3}} f_{5/2} (C)  \makebox[1cm]{\ }
  (\epsilon=1),    \label{lambda0}    
\ee
\be
\alpha \sqrt{2 \pi \rho ( 1 - \rho) N^3} = f_{3/2} (C), 
\label{alpha0} 
\ee
where $f_k (C)$ are defined as 
\bea 
f_{3/2} (C) &=& - \sum_{n=1}^\infty { (-C)^n \over n^{3/2}},  
\label{32}   \\
f_{5/2} (C) &=& - \sum_{n=1}^\infty { (-C)^n \over n^{5/2}},
\eea
for $|C| \leq 1$. To probe the region 
$ \alpha \sqrt{2 \pi \rho ( 1 - \rho) N^3} < f_{3/2} (-1) $, 
$f_k (C)$  are analytically continued as 
\bea
f_{3/2} (C) &=& -4 \sqrt{\pi} 
[ - \ln (-C)]^{1/2}- \sum_{n=1}^\infty { (-C)^n \over n^{3/2}},  
\label{cont32}  \\
f_{5/2} (C) &=& {8 \over 3} \sqrt{\pi} 
[ - \ln (-C)]^{3/2} - \sum_{n=1}^\infty { (-C)^n \over n^{5/2}},
\label{cont52}  
\eea
for $-1 \leq  C < 0$, 
while for $ \alpha \sqrt{2 \pi \rho ( 1 - \rho) N^3} > f_{3/2} (1) $, 
one may use the integral forms 
\bea
f_{3/2} (C) &=& 2 \pi \int_0^\infty ds {\sqrt{s} \over C^{-1} e^{\pi s} +1}, \\
f_{5/2} (C) &=& 2 \pi \int_0^\infty ds \sqrt{s} \ln (1 + C e^{- \pi s}) .
\eea
 
To generalize Eqs.~(\ref{lambda0}) and (\ref{alpha0}) to the case of PASEP 
($0<\epsilon \leq 1$), we limit our attention only to the scaling limit and  
use the results of Kim \cite{dkim}. In \cite{dkim}, 
the low-lying eigenvalues of the asymmetric XXZ chain near the stochastic
line ($\alpha = 0$ in Eq.~(\ref{m})) have been expressed as  perturbative
expansions in $N^{-1/2}$ with a scaling variable, which is essentially
the same as $\alpha N^{-3/2}$, held constant. Therefore the results of
\cite{dkim} applied to the ground state energy (denoted as $E_N^0$ in
\cite{dkim}) can be used immediately to obtain the LDF.

The notations   $q$, $\widetilde{\Delta}$, $s$, $H$, and  $\nu$  used in
\cite{dkim} translate into the present ones as 
$\rho$, $(\cosh \alpha + \epsilon \sinh \alpha)^{-1}$, 
$(\sinh \alpha + \epsilon \cosh \alpha )/(\cosh \alpha 
+ \epsilon \sinh \alpha)$, $(\alpha + \tanh^{-1} \epsilon )/2$, 
and  $ \tanh^{-1} \epsilon$, respectively. 
Using these and taking care of different normalization ($\lambda (\alpha) =
-E_N^0 / \widetilde{\Delta}$), one can rewrite Eq.~(58a) of \cite{dkim}  as
\be
\lambda(\alpha)  =  \epsilon \sum_{m=1}^{\infty} \sum_{k=1}^{m} 
\frac{b_{m,k}}{(1-x_c)^{k+1}}Y_m^0(z) \left( \pi \over N \right)^{m/2},     
\label{lambda}
\ee
and Eq.~(54a) of \cite{dkim} as
\be
\alpha  =  {1 \over \pi} \sum_{m=1}^{\infty} \sum_{k=1}^m b_{m,k}
\frac{(-1)^k }{k\, x_c^k} Y_m^0(z) \left(\frac{\pi}{N}\right)^{(m+2)/2}.
\label{alpha}
\ee
In the above sums, $Y_m^0$ for even $m$ vanishes for the ground state
and only odd-$m$ terms are needed. For $m$ odd, $Y_m^0 (z)$ with real $z$ 
are defined as
\bea
Y_m^0(z) &=&  \frac{m+2iz}{2(m+2)} (-i \sqrt{z+i})^m
              +   \frac{m-2iz}{2(m+2)} (i \sqrt{z-i})^m \nonumber \\
& & + {1 \over {2i}} \int_0^{\infty} dt
\frac{(-i\sqrt{z+i+t})^m - (-i\sqrt{z+i-t})^m}{e^{\pi t} -1} \nonumber  \\
& & - {1 \over {2i}} \int_0^{\infty} dt
\frac{(i\sqrt{z-i+t})^m - (i\sqrt{z-i-t})^m}{e^{\pi t} -1}.
\label{ym0}
\eea
The coefficients $x_c$ and $b_{m,k}$ are recursively
determined order by order in $N^{-1/2}$ from a set of equations, 
as explained in \cite{dkim} and
$x_c = - \rho /(1-\rho)  + O(N^{-5/2})$,  $b_{m,k} = b_{m,k}^0 +
O( N^{-3/2})$. $b_{m,k}$ is the coefficient of $x^m$ 
in the series expansion of $ \left( \sum_{m=1}^\infty a_m x^m \right)^k $ 
where $ a_m = a_m^0 + O(N^{-3/2})$ and 
the first few values of $a_m^0$ needed in this work are given by
\bea
a_1^0 & = & \sqrt{{2 \rho \over (1 - \rho )^3}},    \nonumber  \\
a_2^0 & = & - {2 \over 3} \frac{(1 + \rho)}{(1 - \rho)^2},  \nonumber   \\ 
a_3^0 & = & \sqrt{{2 \rho \over (1 - \rho )^3}} \frac{1 + 11 \rho
             + \rho^2}{18 \rho (1 - \rho )}.  
\label{eq:zerovalue}
\eea
The eigenvalue expression Eq.~(\ref{lambda}) is a power series expansion
in $N^{-1/2}$ with
the scaling variable $\alpha N^{3/2}>0$
and $\epsilon >0$ fixed. (If $\alpha>0$ and finite, the asymmetric XXZ chain
is in the critical phase and hence the ground state energy and the low lying 
excitations possess finite-size corrections analytic in $N^{-1}$.)  
When $\epsilon \rightarrow 0$ with another crossover scaling variable 
$\epsilon \sqrt{N}$ fixed, the infinite series Eq.~(\ref{lambda}) reduces to 
a series in $1/(\epsilon \sqrt{N})$.

Inserting the zeroth order values of $x_c$ and $b_{m,k}$, and keeping only
the leading order terms in Eqs.~(\ref{lambda})  and (\ref{alpha}), 
one then obtains
\bea
\sqrt{{ 2 \pi N^3 \over \rho ( 1 - \rho)}} [ \lambda (\alpha) - \alpha \bar{v}]
& =& \epsilon \left\{ \left( -\frac{4 \pi^2}{3} Y_3^0(z) \right) 
- ( 2 \pi Y_1^0 (z)) \right\} \makebox[1cm]{\ }  (0 < \epsilon \leq 1),   
\label{lambda1}      \\
\alpha  \sqrt{2 \pi \rho(1-\rho)N^3} & = &\left( 2 \pi Y_1^0(z) \right).  
\label{alpha1}
\eea
Here $\bar{v}$ is the exact average current, 
$\epsilon \rho ( 1-\rho) N^2 /(N -1)$. 
The second term on the right-hand side of Eq.~(\ref{lambda1}) appears due to 
our choice of the exact $\bar{v}$ on the left-hand side of Eq.~(\ref{lambda1}).
Except for that, the similarity of Eq.~(\ref{lambda1}) to Eq.~(\ref{lambda0})
is obvious. One simply needs to relate $Y_m^0 (z)$ to $f_k (C)$.  
In Appendix A, we show, by changing the integration contours of 
Eq.~(\ref{ym0}), that $2\pi Y_1^0 (z)$ is indeed 
nothing but a  different form of $f_{3/2} (C)$, 
provided the variables $z$
and $C$ are related by $C = e^{\pi z}$. 
So is $-4 \pi^2 Y_3^0 (z)/3$ of $f_{5/2} (C)$. 
Moreover, we show in Appendix that the 
analytic continuation of Eq.~(\ref{ym0}) to the region Im $z > 1$ naturally
reproduces the analytically continued forms of Eqs.~(\ref{cont32}) and 
(\ref{cont52}).  Therefore, 
the generalization of Eq.~(\ref{lambda0}) to $\epsilon \neq 1$ is achieved by 
a factor $\epsilon$ multiplying the right-hand side of Eq.~(\ref{lambda0}).
Consequently,  by Eqs.~(\ref{leg_f}), (\ref{leg_y}),  (\ref{lambda1}), and 
(\ref{alpha1}), one obtains the LDF in the form 
\be
f(y) \simeq \epsilon \sqrt{\frac{\rho(1-\rho)}{\pi N^3}} H 
\left( \frac{y}{\epsilon \rho (1-\rho)} \right),
\label{scal_LDF}
\ee
where the universal scaling function $H(x)$ is given in the parametric form 
satisfying the relation
\bea
H(x) &=&  {f_{5/2} (C) { f_{3/2}}^{'} (C) - {f_{5/2}}^{'} (C) f_{3/2} (C)  
\over \sqrt{2} f_{3/2}^{'} (C)},   
\label{eq:h} \\  
x &=& {{ f_{5/2}}^{'} (C) - {f_{3/2}}^{'} (C)  \over {f_{3/2}}^{'} (C) }, 
\label{eq:x}    
\eea
with $'$ denoting the derivative with respect to $C$. $H(x)$, as defined here, 
is $H(x+1)$ of \cite{derrleb,derrapp}, 
the difference originating from using exact $\bar{v}$ in Eqs.~(\ref{eq:LDF}) 
and (\ref{lambda1}). 
Thus it has the following asymptotic behaviors 
\be
H(x) \simeq \left\{
\begin{array}{ll} 
-x^2 & \mbox{for} \ |x|\ll 1 ,  \\[2mm] 
-{2 \over 5} \, \sqrt{{3 \over \pi}}  \  x^{5/2}  
&  \mbox{for} \ x \to \infty , \\[2mm] 
- {4 \over 3} \, \sqrt{\pi} \  |x|^{3/2}  &  \mbox{for} \ x \to -\infty . 
\end{array}
\right. 
\label{eq:hproperty}
\ee
Eq.~(\ref{scal_LDF}) is the generalization of the result of 
\cite{derrleb,derrapp}.

\section{Finite-size corrections in discrete dynamics}  
\label{sec:finite}

Finite-size correction is useful in comparing theoretical predictions with 
simulation data. 
In simulations, particle configurations are updated in discrete time steps, 
and to describe such situations, Eqs.~(\ref{masterP}) and (\ref{masterF}) 
should be replaced by their discrete time versions. 
For example, Eq.~(\ref{masterF}) is  replaced by  
\be
F_{\tau + 1} (\sigma; \alpha) - F_{\tau} (\sigma; \alpha) = { 1 \over N} 
\sum_{\sigma'} \langle \sigma | M | \sigma' \rangle F_{\tau} (\sigma'; \alpha),
\label{Fdiff}
\ee
where one update interval is set as $dt = 1/N$ and $t=\tau/N$. 
These difference equations reduce to the continuous time versions, 
Eq.~(\ref{masterP}) and Eq.~(\ref{masterF}), in the limit $N \to \infty$. 
Thus the leading terms in $N$ of all quantities are the same in both versions. 
However, there appear differences in the finite-size corrections and we
work in the discrete version.
Using Eq.~(\ref{Fdiff}), Eq.~(\ref{Ylambda}) is then modified to  
$\langle e^{\alpha Y_t} \rangle = \sum_{\sigma} F_t (\sigma; \alpha) 
\sim e^{ \mu(\alpha) t}$, where
\be 
\mu(\alpha) = N \ln \left( 1 + {\lambda(\alpha) \over N} \right).
\label{Ymu}
\ee
Therefore the LDF 
is the Legendre transformation of $\mu(\alpha) - \alpha \bar{v}$.  

From Eqs.~(\ref{lambda}), (\ref{alpha}), and (\ref{Ymu}), 
$\mu(\alpha) - \alpha \bar{v}$ is written as, 
including its next leading term, 
\be
\sqrt{2 \pi N^3 \over \rho ( 1 - \rho)} [ \mu(\alpha) - \alpha \bar{v} ] 
\simeq   \epsilon \{ f_{5/2} (C) - f_{3/2} (C) \}  
- { \epsilon^2 \over 2 \sqrt{2} }  \sqrt{ \rho ( 1 - \rho) \over  \pi N } 
f_{3/2} (C)^2.
\label{mu1}
\ee
The last term on the right-hand side of Eq.~(\ref{mu1}) arises from the 
first nonlinear term in the expansion  
$\mu(\alpha) = \lambda(\alpha) - \lambda(\alpha)^2 / 2N  + \cdots$. 
The leading correction term in $\mu(\alpha) - \alpha \bar{v}$ is
of order $N^{-1/2}$, 
while that in $\lambda(\alpha) - \alpha \bar{v}$ is of order $N^{-1}$. 
Since the leading correction to $\alpha \sqrt{2 \pi \rho(1-\rho)N^3}$ is also 
of order $N^{-1}$, using Eq.~(\ref{alpha1}) and  Eq.~(\ref{mu1}), 
one finds that
\be
f(y) = \epsilon \sqrt{\rho(1-\rho) \over \pi N^3} \left[ 
H\left( { y \over \epsilon \rho ( 1 - \rho) } \right) + 
\epsilon \sqrt{ { \rho ( 1 - \rho ) \over   \pi N  } }  H_1 \left( 
{ y \over \epsilon \rho ( 1 - \rho) } \right) + O(N^{-1}) \right],
\label{finite_LDF}
\ee
where $H_1 (x)$ is determined from Eq.~(\ref{eq:x}) and
\be
H_1 (x) = - { f_{3/2} (C)^2 \over 4}.
\ee
The correction term shows dependence on the particle density and 
the asymmetry parameter, and hence  is not  universal. 
The asymptotic behaviors of $H_1 (x)$ are 
\be
H_1 (x) \simeq \left\{
\begin{array}{ll}
- 2 x^2   & \mbox{for } |x|\ll 1,    \\[2mm]
- 3 x^3 /(2 \pi)   & \mbox{for }   x \to \infty ,    \\[2mm]
- 2 \pi |x|   & \mbox{for } x \to -\infty.
\end{array}
\right.
\label{h1prop}
\ee

Another quantity of interest concerning the finite-size correction is 
the cumulant ratio considered in \cite{derrapp}. 
It is defined as
\be
\lim_{t \rightarrow \infty} R_t = \lim_{t \rightarrow \infty} { {\langle Y_t^3
\rangle}_c^2 \over \langle Y_t^2 \rangle_c \langle Y_t^4 \rangle_c },
\ee
where $ \langle Y_t^n \rangle_c $ are the cumulants of $Y_t$ 
and are evaluated from
\be
\lim_{t \rightarrow \infty} {\langle Y_t^n \rangle_c \over t} = \left. 
{d^n \mu(\alpha) \over d \alpha^n}\right|_{\alpha = 0}.
\ee
Using Eqs.~(\ref{alpha1}) and (\ref{mu1}), we find  
\bea
\lim_{t \rightarrow \infty} {\langle Y_t^2 \rangle_c \over t} &=& 
\epsilon N^{3/2} [ \rho ( 1-\rho)]^{3/2} { \sqrt{ \pi}  \over 2 }
\left[ 1 - 2 \epsilon \sqrt{\rho ( 1 - \rho) \over \pi N} + O(N^{-1})
\right], \nonumber \\
\lim_{t \rightarrow \infty } {\langle Y_t^3 \rangle_c \over t} &=& \epsilon N^3 
[ \rho ( 1 - \rho )]^2 \pi \left( {3 \over 2} - { 8 \sqrt{3} \over 9 } \right) 
\left[ 1 + O(N^{-1}) \right],  \nonumber  \\
\lim_{t \rightarrow \infty } {\langle Y_t^4 \rangle_c \over t} &=& 
\epsilon N^{9/2} 
[ \rho(1 - \rho)]^{5/2} \pi^{3/2} \left( {15 \over 2 } + { 9 \sqrt{2} \over 2} 
- 8 \sqrt{3} \right) \left[ 1 + O(N^{-1}) \right].
\eea
Therefore the cumulant ratio has an $O(N^{-1/2})$ correction term;  
\be
\lim_{t \rightarrow \infty } R_t
=  2
{ \left( { 3 \over 2}- { 8 \sqrt{3} \over 9 } \right)^2 \over 
\left( { 15 \over 2} + { 9 \sqrt{2} \over 2 }- 8 \sqrt{3} \right) } 
\left( 1 +  2 \epsilon \sqrt{ { \rho ( 1 - \rho) \over \pi N } } 
+ O(N^{-1}) \right).
\label{finite_cum}
\ee

We note in passing that in the continuous time version, our method shows
\be
\lim_{t \rightarrow \infty} {\langle Y_t^2 \rangle_c \over t} =
\epsilon N^{3/2} [ \rho ( 1-\rho)]^{3/2} { \sqrt{ \pi}  \over 2 }
\left[ 1 + {1 + 11 \rho - 11 \rho^2 \over 8 \rho (1 - \rho) N} + O(N^{-3/2})
\right].
\label{Y2cont}
\ee
This is in exact agreement with the expansion derived, 
with the help of Stirling's formula, from Eq.~(6) of Derrida and Mallick 
\cite{derrmal}.

\section{Discussions}
\label{sec:discussion}

The main results of this paper are Eqs.~(\ref{scal_LDF}), (\ref{finite_LDF}), 
and (\ref{finite_cum}). 
The universal scaling function of the LDF, $H(x)$, first defined in 
\cite{derrleb,derrapp} for the TASEP, is reproduced for the PASEP in 
Eq.~(\ref{scal_LDF}). 
The only change in this generalization is the modification of the scaling
variable by a simple factor $\epsilon$, the asymmetry parameter.
Physically, this is equivalent to a rescaling of time by $\epsilon$.
Non-trivial $\epsilon$-dependence of $\lambda(\alpha)$ appears only
in higher orders of $N^{-1/2}$ in Eq.~(\ref{lambda}).
To compare analytic results with simulation data, the finite-size correction
terms in the discrete time dynamics are important. 
They are derived for the LDF and the cumulant ratio in Eq.~(\ref{finite_LDF}) 
and (\ref{finite_cum}), respectively.
One sees that the finite-size corrections in the discrete time dynamics are of
$O(N^{-1/2})$. Also they depend on $\rho$ and $\epsilon$ explicitly in 
both versions implying that they are non-universal.

Instead of the statistics of $Y_t$, the total displacement, 
one could have asked for the statistics of the displacement across one bond.
In this case, one has to deal with an asymmetric $XXZ$ chain with 
a twisted boundary condition, 
$\sigma_{N+1}^{\pm} = e^{\mp \alpha} \sigma_1^{\pm}$ and
$\sigma_{N+1}^z = \sigma_1^z$,
and analysis similar to that presented here can be carried out \cite{dkim2}.
In particular, if $J_t$ is the displacement across the $N$-th bond, 
one can show that 
$\langle e^{\alpha J_t} \rangle \sim \langle e^{\alpha Y_t /N} \rangle 
\sim e^{\lambda(\alpha /N) t}$ (in the continuous time notation).
Therefore, the LDF and the long time behaviors of the cumulants of $J_t$ 
are the same as those of $Y_t /N$.
This is why Eq.~(\ref{Y2cont}) agrees with the result of \cite{derrmal}
where $\lim_{t \to \infty} \langle J_t^2 \rangle_c /t$ is obtained.
However, $\langle {J_t}^2 \rangle_c - \langle (Y_t /N)^2 \rangle_c$,
the surface width in the growth model language, saturates to a finite value
of $O(N)$ as $t \to \infty$.

\acknowledgements
We thank B. Derrida and J.M. Kim for helpful discussions. 
This work is supported by the Korea Research Foundation grant 
1998-015-D00055 and also 
by the Center for Theoretical Physics, Seoul National University. 

\appendix

\section{Properties of $Y_{\lowercase{m}}^0 (\lowercase{z})$}

In this Appendix, we show  the equivalence of $f_{ 1 + m/2} (C)$ and 
$ Y_m^0 (z) \  ( m = 1 , 3 , 5 , \ldots )$. 
The former will be defined later extending  the definitions of $f_{3/2} (C)$ 
and $f_{5/2} (C)$, and the latter is defined in Eq.~(\ref{ym0}). 
We take Eq.~(\ref{ym0}) as defining $Y_m^0 (z)$ for any complex $z$. 

\subsection{Simple form of $Y_1^0 (z) $ }

\begin{figure}[t]
\epsfxsize=150mm \epsfysize=60mm \centerline{\epsfbox{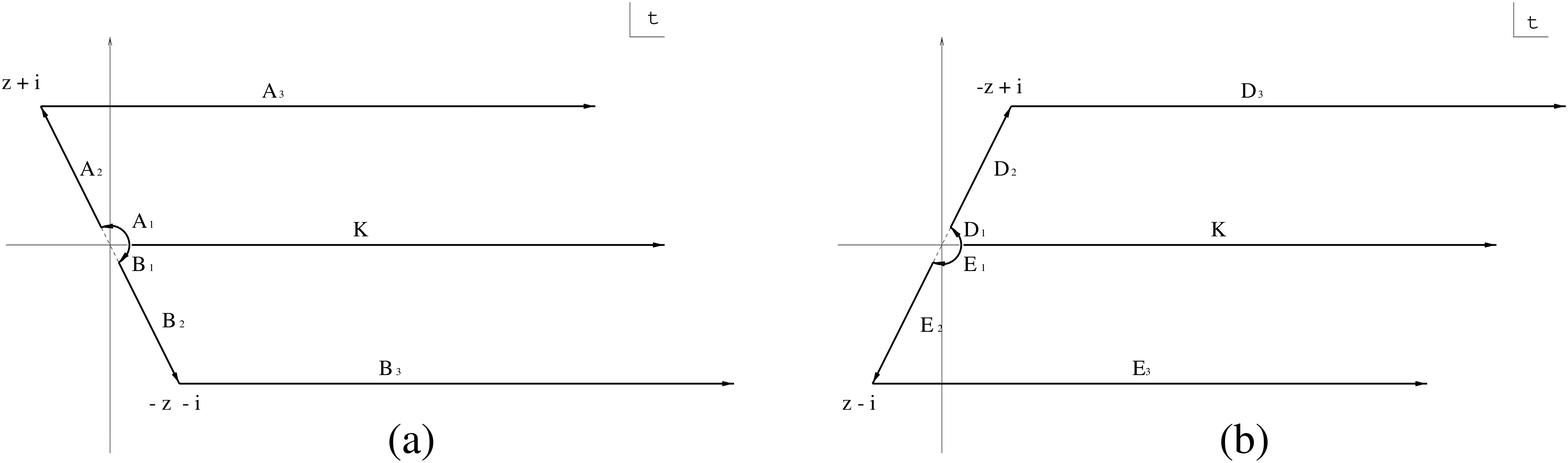}} \vspace{1pc}
\caption{ Contours for $I_1$ (a) and $I_2$ (b). Re $z$ is assumed to be negative
and $|{\rm Im} \ z| < 1$. The semicircles about the origin have
the small radii $\delta$, which are set to be zero in the last step.}
\label{fig:contour1}
\end{figure}

We first pay attention to $Y_1^0 (z)$ 
since $Y_m^0 (z)$ ($m = 3, 5, \ldots$)  can be evaluated from 
$Y_1^0 (z) $ through the recursion relation,
$d Y_{m+2}^0 (z) /dz   =  -(m + 2) Y_m^0 (z) /2$ \cite{dkim}.
Eq.~(\ref{ym0}) for $m = 1$ is written as
\bea
2 Y_1^0 (z)  & = & -i (z + i)^{1/2} + { 2 \over 3 } ( z+ i )^{3/2} + 
\int_0^\infty  dt {(z + i - t)^{1/2} - (z + i + t)^{1/2} \over e^{\pi t} -1} 
\nonumber   \\
&  &  + i (z - i)^{1/2} + { 2 \over 3 } ( z- i )^{3/2} + 
\int_0^\infty  dt {(z - i - t)^{1/2} - (z - i + t)^{1/2} \over e^{\pi t} -1}.   
\label{eq:yrecursion} 
\eea
Suppose $|{\rm Im} \ z| < 1$ and 
let $I_1$ and $I_2$ be the first and the second integrals
in Eq.~(\ref{eq:yrecursion}), respectively.
The two integrations are over the positive real axis of the complex-$t$ 
plane, denoted by $K$ in Fig.~\ref{fig:contour1}. 

Our method is to deform the contours in the complex-$t$ plane such that 
only simple integrals remain and the additive terms cancel out. 
Each integral has two terms. 
For the first term of $I_1$, $K$ is deformed to $A_1 + A_2 + A_3$ as 
shown in Fig.~\ref{fig:contour1}(a) 
while for the second term of $I_1$, 
to $B_1 + B_2 + B_3$ in  Fig.~\ref{fig:contour1}(a).
Similarly, $K$ is deformed to $E_1 + E_2 + E_3 $ and $D_1 + D_2 + D_3$
for the first and second terms of $I_2$, respectively,
as shown in  Fig.~\ref{fig:contour1}(b).
We then have
\bea
I_1 & = & \int_{A_1 + A_2 + A_3}  dt 
{(z + i - t)^{1/2} \over e^{\pi t} -1} 
- \int_{B_1 + B_2 + B_3}  dt 
{(z + i + t)^{1/2} \over e^{\pi t} -1}   \nonumber   \\
& = & \int_0^\theta d \theta' {i (z + i)^{1/2} \over \pi} 
- \int_0^{- (\pi - \theta)} d \theta' {i (z + i)^{1/2} \over \pi}
\nonumber  \\
& & + \int_0^1 d \xi {(z + i)^{3/2} (1 - \xi)^{1/2} \over 
e^{ \pi (z + i) \xi} - 1} 
- \int_0^1 d \xi {- (z + i)^{3/2} (1 - \xi )^{1/2} \over 
e^{- \pi (z + i) \xi } - 1}  \nonumber   \\ 
&  &  + \int_0^{\infty} ds 
{(- s)^{1/2} \over e^{\pi (z + i)} e^{\pi s} - 1} 
- \int_0^{\infty}  ds {s^{1/2} \over e^{- \pi (z + i)} e^{\pi s} - 1}
\nonumber   \\
& = & i (z + i)^{1/2} - {2 \over 3} (z + i)^{3/2} 
+ \int_0^{\infty} ds {(- s)^{1/2} \over e^{\pi (z + i)} e^{ \pi s } - 1} 
- \int_0^{\infty} ds {s^{1/2} \over e^{- \pi (z + i)} e^{ \pi s } - 1}, 
\label{eq:I1}
\eea
where $\theta =$ Arg($z + i$). Similarly,
\bea
I_2 & = & \int_{E_1 + E_2 + E_3}  dt 
{(z - i - t)^{1/2} \over e^{\pi t} -1} 
- \int_{D_1 + D_2 + D_3}  dt 
{(z - i + t)^{1/2} \over e^{\pi t} -1}   \nonumber   \\
& = & \int_0^{- (\pi - \theta)}  d \theta' {i (z - i)^{1/2} \over \pi} 
- \int_0^{\theta}  d \theta' { i (z - i)^{1/2} \over \pi } \nonumber   \\ 
& & + \int_0^1 d \xi {(z - i)^{3/2} (1 - \xi)^{1/2} \over 
e^{ \pi (z - i) \xi} - 1}
- \int_0^1 d \xi { - (z - i)^{3/2} (1 - \xi)^{1/2} \over 
e^{- \pi (z - i) \xi} - 1}  \nonumber   \\ 
&  &   + \int_0^{\infty} ds 
{(- s)^{1/2} \over e^{\pi (z - i)} e^{\pi s} - 1} 
- \int_0^{\infty}  ds {s^{1/2} \over e^{- \pi (z - i)} e^{\pi s} - 1}    
\nonumber   \\ 
& = & - i (z - i)^{1/2} - {2 \over 3} (z - i)^{3/2} 
+ \int_0^{\infty} ds {(- s)^{1/2} \over e^{ \pi (z - i)} e^{\pi s} - 1} 
- \int_0^{\infty}  ds {s^{1/2} \over e^{- \pi (z - i)} e^{\pi s} - 1},  
\label{eq:I2}
\eea
with $\theta =$ Arg($-z + i$).

We note that the branch cuts for the square-root functions 
in $I_1 $ and $ I_2 $ are in the opposite directions  \cite{dkim}, 
so the two integrals having the factor $(- s)^{1/2} $ in their integrands 
cancel out when $I_1$ and $I_2$ are added.
Therefore we arrive at the conclusion that 
\be
Y_1^0 (z) = \int_0^{\infty}  ds {s^{1/2} \over e^{- \pi z} e^{\pi s} + 1} 
\makebox[1cm]{\ }
( |{\rm Im} \ z| < 1 ).
\label{simy1}
\ee

\begin{figure}[t]
\epsfxsize=150mm \epsfysize=60mm \centerline{\epsfbox{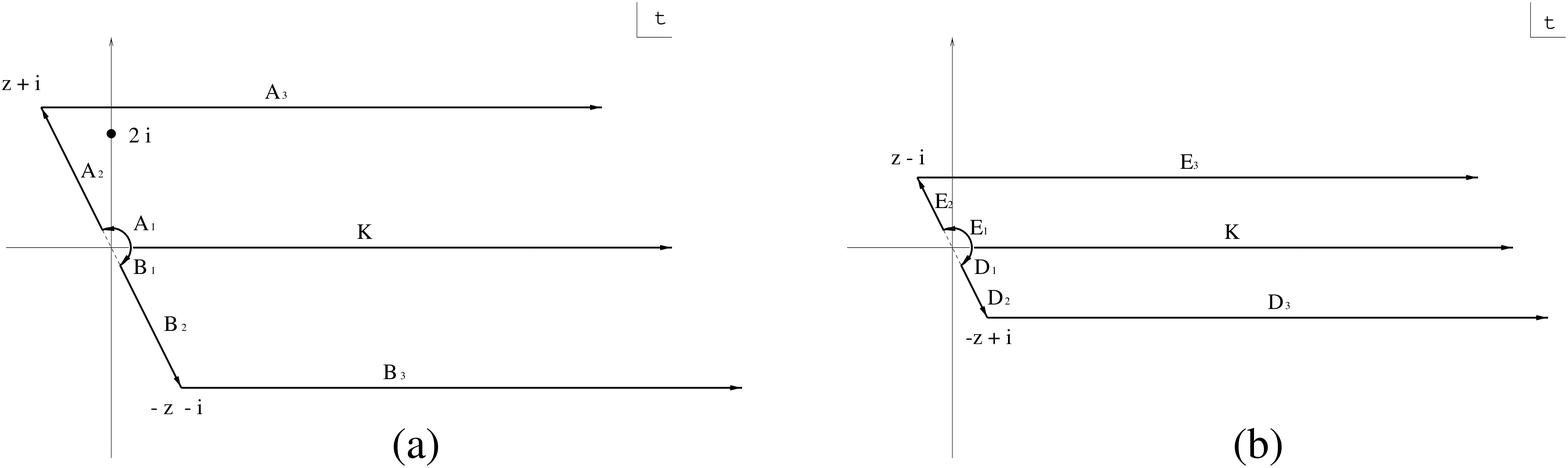}} \vspace{0.5pc}
\caption{ Contours for $I_1$ (a)  and $I_2$ (b).
${\rm Re} \ z $ is assumed to be negative and $ 1 < {\rm Im} \ z < 3 $.
Compared with Fig.~\ref{fig:contour1}, a pole is placed inside the contour
for $I_1$ and the integration over the semicircle about the origin
makes a different(sign-changed) value in $I_2 $.}
\label{fig:contour2}
\end{figure}

Next, consider the region $1 < | {\rm Im} \ z | < 3$.  
If $1 < {\rm Im} \ z < 3$, the contours shown in Fig.~\ref{fig:contour1} 
change to those shown in Fig.~\ref{fig:contour2}.
Compared with Fig.~\ref{fig:contour1}, 
a pole at $t=2i$ is placed inside the contour for $I_1$ and
the direction of the integration over the semicircle about the origin
is reversed for $I_2$. 
These changes produce extra contributions $2i (z - i)^{1/2}$
for both $I_1$ and $I_2$.
Therefore, we obtain 
\bea
Y_1^0 (z) &=& 2 i (z - i)^{1/2} 
+ \int_0^{\infty}  ds {s^{1/2} \over e^{- \pi z} e^{\pi s} + 1} 
\nonumber   \\
&=& - 2 ( - z + i )^{1/2} 
+ \int_0^{\infty}  ds {s^{1/2} \over e^{- \pi z} e^{\pi s} + 1}
\makebox[1cm]{\ } (1 < {\rm Im} \ z < 3).
\label{conpy1}
\eea
Similarly, for $- 3 < {\rm Im} \ z < - 1$, we find
\bea
Y_1^0 (z) &=& - 2 i (z + i)^{1/2} 
+ \int_0^{\infty}  ds {s^{1/2} \over e^{- \pi z} e^{\pi s} + 1} 
\nonumber \\
&=& - 2 (- z - i)^{1/2} 
+ \int_0^{\infty}  ds {s^{1/2} \over e^{- \pi z} e^{\pi s} + 1}
\makebox[1cm]{\ } (-3 < {\rm Im} \ z < -1).
\label{conmy1}
\eea

When $|{\rm Im} \ z | $ increases further, 
more and more poles are placed inside the contours for $I_1 $ and $ I_2 $, 
and corresponding residues should be added.
But recalling that  $Y_m^0 (z)$ is used to express  
real $\alpha $ and $\lambda(\alpha)$, 
Eqs.~(\ref{simy1}) and (\ref{conpy1}) are sufficient for our purpose.

\subsection{Relation between  $f_{1 + m/2} (C) $ and $ Y_m^0 (z) $ }

We now make the identification $C =  e^{\pi z}$ with $C$ real.
$C > 0$ if $z$ is real, and $-1 < C < 0$ if $z = -x + i^-$ with $x > 0$.
In both cases, we have, from Eq.~(\ref{simy1}),
\be
Y_1^0 (z)  =  \int_0^{\infty}  ds {  s^{1/2} \over C^{-1}  e^{ \pi s } + 1 }.
\label{Cy1}
\ee
Eq.~(\ref{Cy1}) admits a series expansion 
\be
Y_1^0 (z) = {1 \over 2 \pi} \sum_{n = 1}^{\infty}  
{(-1)^{n + 1} C^n \over n^{3/2}},
\ee
when $|C| < 1$. The second branch in the region $-1 < C < 0$ is obtained
if  $z = -x + i^+$ with $x > 0$. In this case, using Eq.~(\ref{conpy1}),
\bea
Y_1^0 (z)  &=& - {2 \over \sqrt{\pi}} [- \ln(- C)]^{1/2}
+ \int_0^{\infty}  ds {s^{1/2} \over C^{-1}  e^{\pi s} + 1} \nonumber  \\
&=& -  {2 \over \sqrt{\pi}}  [- \ln(- C)]^{1/2}
+ {1 \over 2 \pi} \sum_{n = 1}^{\infty}  {(-1)^{n + 1} C^n \over n^{3/2}}.  
\label{addCy1}
\eea
Comparing Eqs.~(\ref{Cy1}) and (\ref{addCy1}) with Eqs.~(\ref{32}) and 
(\ref{cont32}), we have
\be
f_{3/2} (C) = 2 \pi Y_1^0 (z),
\label{f32y1}
\ee
provided $C = e^{\pi z}$, the two branches of $-1 < C < 0$ corresponding to
$|{\rm Im} \ z| = 1^-$ and $|{\rm Im} \ z| = 1^+$, respectively.

Next, we define $f_{1 + m/2} (C) \ (m = 1, 3, 5, \ldots)$ by
Eq.~(\ref{f32y1}) and the recursion relation
\be
{d \over  d(\ln C)} f_{1 + (m+2)/2} (C) = f_{1 + m/2} (C),
\ee
with the initial condition $f_{1 + m/2} (0) = 0$.
Then $f_{1 + m/2} (C)$ takes the form   
\be
f_{1 + m/2} (C) =  \sum_{n = 1}^{\infty}  {(-1)^{n + 1} C^n \over n^{1 + m/2}},
\ee
in the first branch and 
\be
f_{1 + m/2} (C) =  (-1)^{ (1 + m)/2} {(1/2)! \over (m/ 2)!} 4 \sqrt{\pi} 
[- \ln(- C)]^{m/2}   
+ \sum_{n = 1}^{\infty}  {(-1)^{n + 1} C^n \over n^{1 + m/2}},
\ee
in the second branch. 
Comparison of the recursion relations of $Y_m^0 (z)$ and $f_{1 + m/2} (C)$  
then leads to the identification
\be
f_{1 + m/2} (C) = (-\pi)^{(m-1)/2} 2 \pi { (1/2)! \over (m/2)!} Y_m^0 (z).
\ee
For example, 
$f_{3/2} (C) = 2 \pi Y_1^0 (z), \  
f_{5/2} (C) = - 4 \pi^2 Y_3^0 (z) /3, \    
f_{7/2} (C) = 8 \pi^3 Y_5^0 (z) /15$, etc.
  


\end{document}